\documentclass[conference]{IEEEtran}
\IEEEoverridecommandlockouts
\usepackage{cite}
\usepackage{amsmath,amssymb,amsfonts}
\usepackage{algorithmic}
\usepackage{graphicx}
\usepackage{subcaption}
\usepackage{textcomp}
\usepackage{fancyhdr}
\def\BibTeX{{\rm B\kern-.05em{\sc i\kern-.025em b}\kern-.08em
    T\kern-.1667em\lower.7ex\hbox{E}\kern-.125emX}}
\begin{document}

\title{Emotion Detection and Analysis on Social Media\textsuperscript{\fontsize{12}{20}\selectfont 1}}

%\author{
%\IEEEauthorblockN{Bharat Gaind, Varun Syal, Sneha Padgalwar}
%\IEEEauthorblockA{\textit{Department Of Computer Science And Engineering}\\
%\textit{Indian Institute Of Technology, Roorkee, India}\\
%Email: \{bharatgaind234, varunsyal1994, sneha.padgalwar\}@gmail.com}
%}

\author{
\IEEEauthorblockN{Bharat Gaind}
\IEEEauthorblockA{\textit{CSE Department}\\
\textit{IIT Roorkee}\\
Roorkee, India \\
bharatgaind234@gmail.com}
\and
\IEEEauthorblockN{Varun Syal}
\IEEEauthorblockA{\textit{CSE Department}\\
\textit{IIT Roorkee}\\
Roorkee, India \\
varunsyal1994@gmail.com}
\and
\IEEEauthorblockN{Sneha Padgalwar}
\IEEEauthorblockA{\textit{CSE Department}\\
\textit{IIT Roorkee}\\
Roorkee, India \\
sneha.padgalwar@gmail.com}
}

\thispagestyle{fancy}\fancyhf{}
\renewcommand{\thispagestyle}[1]{}\date{}
\renewcommand{\headrulewidth}{0pt}
\renewcommand{\footrulewidth}{0.1pt}

\lfoot{\textsuperscript{1}\footnotesize \fontsize{9}{9.6} \selectfont Published in the Global Journal of Engineering Science and Researches (ISSN 2348 - 8034, Pg. 78-89) after getting accepted in the International Conference on Recent Trends In Computational Engineering and Technologies (ICRTCET'18), May 17-18, 2018, Bengaluru, India. (Journal Link: http://www.gjesr.com/ICRTCET-18.html) }

\maketitle

\begin{abstract}
In this paper, we address the problem of detection, classification and quantification of emotions of text in any form. We consider English text collected from social media like Twitter, which can provide information having utility in a variety of ways, especially opinion mining. Social media like Twitter and Facebook is full of emotions, feelings and opinions of people all over the world. However, analyzing and classifying text on the basis of emotions is a big challenge and can be considered as an advanced form of Sentiment Analysis. This paper proposes a method to classify text into six different \emph{Emotion-Categories}: Happiness, Sadness, Fear, Anger, Surprise and Disgust. In our model, we use two different approaches and combine them to effectively extract these emotions from text. The first approach is based on Natural Language Processing, and uses several textual features like emoticons, degree words and negations, Parts Of Speech and other grammatical analysis. The second approach is based on Machine Learning classification algorithms. We have also successfully devised a method to automate the creation of the training-set itself, so as to eliminate the need of manual annotation of large datasets. Moreover, we have managed to create a large bag of emotional words, along with their emotion-intensities. On testing, it is shown that our model provides significant accuracy in classifying tweets taken from Twitter.
\end{abstract}

\begin{IEEEkeywords}
Sentiment Analysis, Machine learning, Data mining, Natural Language Processing
\end{IEEEkeywords}

\section{Introduction}
Emotions are described as intense feelings that are directed at something or someone in response to internal or external events having a particular significance for the individual. And the internet, today, has become a key medium through which people express their emotions, feelings and opinions. Every event, news or activity around the world, is shared, discussed, posted and commented on social media, by millions of people. Eg. \emph{``The Syria chemical attacks break my heart!! :'("} or \emph{``Delicious dinner at Copper Chimney! :D"} or ``OMG! That is so scary!". Capturing these emotions in text, especially those posted or circulated on social media, can be a source of precious information, which can be used to study how different people react to different situations and events.

Business analysts can use this information to track feelings and opinions of people with respect to their products. The problem with most of the Sentiment Analysis that is done today is that the analysis only informs whether the public reaction is positive or negative but fails to describe the exact feelings of the customers and the intensity of their reaction. With our emotional analysis, they can have a more profound analysis of their markets than the naive 2-way Sentiment Analysis, which itself has turned their businesses more profitable. Business leaders can analyse the holistic view of people in response to their actions or events and work accordingly. Also, health-analysts can study the mood swings of individuals or masses at different times of the day or in response to certain events. It can also be used to formulate the mental or emotional state of an individual, studying his/her activity over a period of time, and possibly detect depression risks.

There are plenty of research works that have focussed on Sentiment Analysis and provide a 2-way classification of text. But few have actually focussed on mining emotions from text. However, machine analysis of text to classify and score it on the basis of emotions poses the following challenges :
\begin{itemize}
\item Instead of the usual two categories in Sentiment Analysis, there are six \emph{Emotion-Categories} in which we need to classify the tweets.
\item Lack of manually annotated data to train classifiers to label data into six categories.
\item Unavailability of a comprehensive bag of \emph{Emotion-words} labeled and scored according to \emph{Emotion-Categories} (Happiness, Sadness, etc.) and their intensities, that can be used to detect \emph{Emotion-words} in text.
\end{itemize}
In order to address the aforementioned challenges, it was important to devise a system that could generate a good and reliable training-set for the classifier, a labeled bag of words, and an algorithm that could not only detect emotions, but also score and label the tweets according to those emotions.

A lot of research has been done on classifying comments, opinions, movie/product reviews, ratings, recommendations and other forms of online expression into positive or negative sentiments. Emotions have also been studied, but in a limited extent, such as by asking specific questions and judging on the basis of replies, or an analysis done only on short one-lined headlines or a few others \cite{balabantaray2012multi}, \cite{alm2005emotions}, \cite{pearl2010identifying}, all of which depended on the manual annotation of the training dataset of a small size and limited scope.

In this paper, we propose a method to classify and quantify tweets according to six standard emotions suggested by Paul Ekman \cite{ekman1992argument}. Here, we base our analysis on tweets posted on Twitter, but it can be easily extended to any kind of text whether it is one lined headlines, messages and posts on social media or larger chunks of writings, because of automatic development of our training set. Our main contributions are listed below:
\begin{itemize}
\item We have developed a system that could score and label any piece of text, especially tweets and posts on social media according to six \emph{Emotion-Categories}: Happiness, Sadness, Fear, Surprise, Anger and Disgust along with their intensity scores, making use of its textual features, a variety of NLP tools and standard Machine Learning classifiers.
\item Another significant contribution is that we have successfully devised a system that could automatically (without any manual effort) build an efficient training set for our ML Classifiers, consisting of a large enough set of labeled tweets from all \emph{Emotion-Categories}.
\item We have created a large bag of words in English, that consists of words expressing a particular emotion along with the intensity of that emotion.
\item We were able to achieve an accuracy of about 91.7\% and 85.4\% using J48 and SMO classifiers respectively using the training set we built.
\end{itemize}

\section{Related Work}

In the recent past, with the rise of social media such as blogs and social networks, a lot of interest has been fueled in Sentiment Analysis. Lately, a lot of research has been done on classifying comments, opinions, movie/product reviews, ratings, recommendations and other forms of online expressions into positive or negative sentiments. Earlier research involved manually annotated corpus of limited size to classify the emotions. Wiebe et al. \cite{wiebe2005annotating} worked on the manual annotation of emotions, opinions and sentiments in a sentence corpus (of size 10,000) of news articles. Segundo et al. \cite{segundo2008language} also studies the presence of emotions in text, and is a functional theory of the language used for expressing attitudes, judgments and emotions \cite{pearl2010identifying}. This paper deals explicitly with emotions, which none of the aforementioned works do. There was, however, one research work \cite{strapparava2008learning}, that classified text into six \emph{Emotion-Categories}, but that was only limited to classification of news headlines, and the training set used was created manually. We, on the other hand, have developed a system, which classifies text in any form (eg. news, tweets, or narrative) and uses a training set, which is generated automatically. This saved a lot of effort. Another similar work was of Carlo and Rada \cite{strapparava2007semeval} who did a similar classification on news headlines but even they used manually annotated corpus for their classification. There is one research work \cite{wang2012harnessing}, that explores the possibility of creating automatic training datasets, like we do and also tries to find out if creating large emotion datasets can increase the emotion-detection accuracy in tweets. Some aspects of their work are similar to ours, but the emotion-detection accuracy we have achieved is much higher. Another similar work is \cite{aman2007identifying}, but again, our accuracy is much higher. A recent work \cite{liu2017applications} explores the possibility of predicting future stock returns based on tweets related to presidential elections and NASDAQ-100 companies. Another recent work \cite{stojanovski2018emotion} uses convolutional neural network architecture for emotion identification in Twitter messages. Their approach uses unsupervised learning, whereas we use supervised learning and their accuracy of 55.77\% is much lower than ours.

\section{Data Sets}
This section describes the various data sets used, such as \emph{Tweets Set}, \emph{Emotion-Words Set (EWS)}, \emph{Degree-Words Set} and \emph{Location-Areas Set}.

\subsection{Tweets Set}
We use Tweepy \cite{tweepy} to collect tweets, which is a Python library for accessing the Twitter API. It takes as input various parameters, such as coordinates, radius, etc., and after removal of duplicates, links, hashtags, and words in other languages (besides English) from these tweets, stores the tweet-ids, text and location of the most recent ones in the database. This provided us with a list of tweets from various locations across the country. Eg. The \textbf{\emph{Tweets Set}}, we created for Delhi has about 10,000 entries. Another way we used Tweepy is by feeding it a twitter-username (of a user) as an input to store all the tweets of that user (till date), in our database.

\renewcommand{\arraystretch}{2}
\begin{table}[h]
\centering
\caption{An example of \emph{Emotion-Words Set (EWS)}}
\label{ews}
\begin{tabular}{|l|l|l|}
\hline
\textbf{Word/ Emoticon} & \textbf{Emotion-Category} & \textbf{Intensity-Category} \\ \hline
:O                      & SURPRISE                  & STRONG                      \\ \hline
Repugnance              & DISGUST                   & STRONG                      \\ \hline
Delighted               & HAPPINESS                 & MEDIUM                      \\ \hline
Afraid                  & FEAR                      & STRONG                      \\ \hline
:(                      & SADNESS                   & STRONG                      \\ \hline
Irritated               & ANGER                     & STRONG                      \\ \hline
Lucky                   & HAPPINESS                 & LIGHT                       \\ \hline
\end{tabular}
\end{table}

\begin{figure*}[]
  \begin{center}
  \includegraphics[scale=0.25]{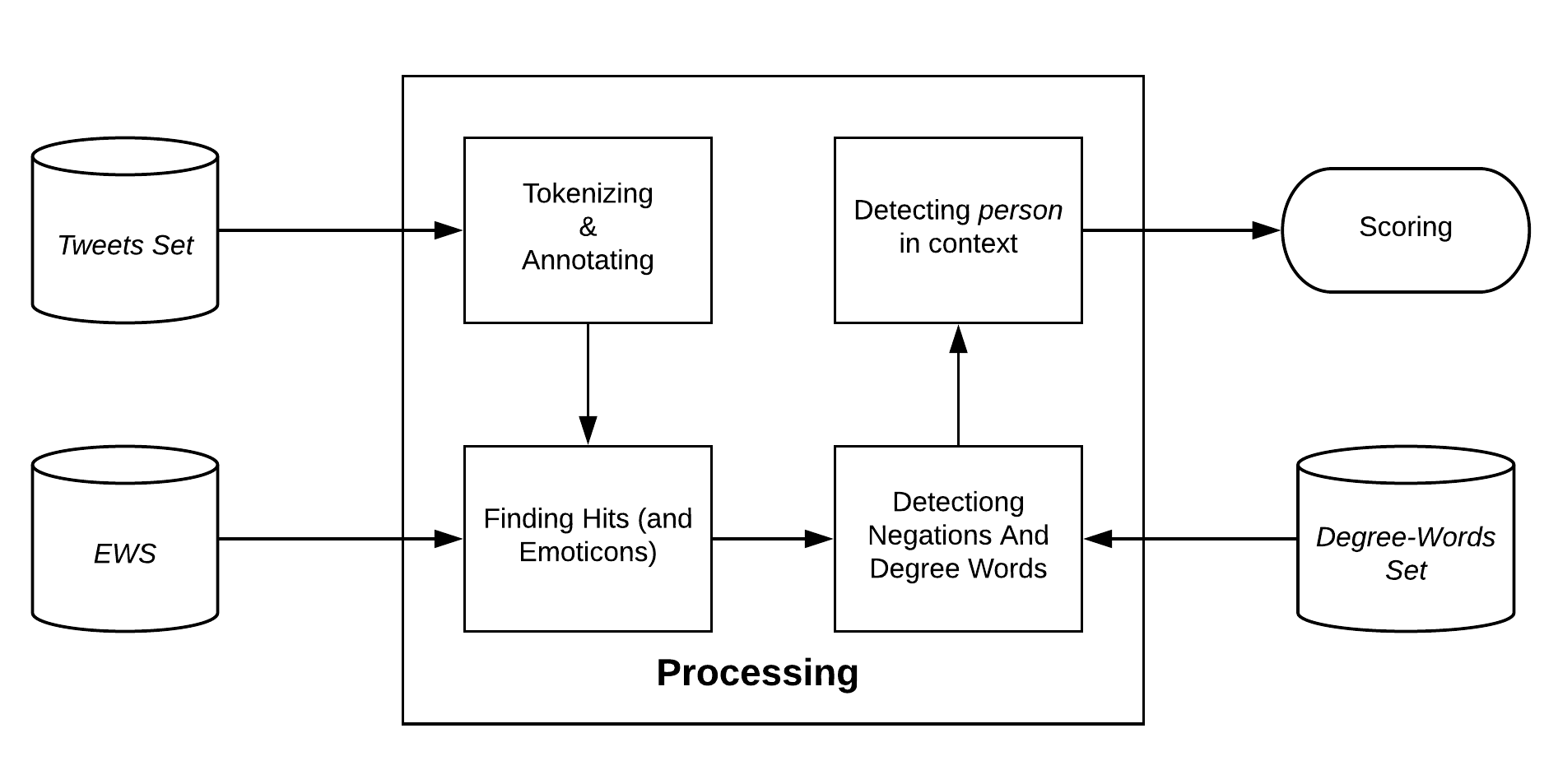}
  \caption{An overview of the first approach}
  \label{first_approach}
  \end{center}
\end{figure*}

\subsection{Emotion-Words Set (EWS)}
A proper selection of relevant and commonly-used \emph{Emotion-words} is one of the most essential and indispensable aspects in the emotional quantification of a sentence. We have created a high-quality accurate bag of words, called \textbf{\emph{Emotion-Words Set (EWS)}}, of around 1500 words. It has been developed by recursively searching (Depth First Search) the synonyms of the 6 basic \emph{\textbf{Emotion-Categories} (HAPPINESS, SADNESS, ANGER, SURPRISE, FEAR, DISGUST)} in a thesaurus \cite{thesaurus}, up to two levels. Each of these words was then manually labeled to one of the three \emph{\textbf{Intensity-Categories} (STRONG, MEDIUM, LIGHT)}, as shown in Table \ref{ews}.

\subsection{Degree-Words Set}
\textbf{\emph{Degree-Words Set}} is a set of about 50 degree words, that are used to strengthen or weaken the intensity of emotions in a sentence. Eg. \emph{``too happy"} and \emph{``hardly happy"} have two different meanings, almost opposite to each other. In this set, each word has an associated \textbf{\emph{Degree-Intensity}} stored with it; \emph{\textbf{H}} meaning High (having a high multiplying effect), \emph{\textbf{N}} meaning Negation (having an opposing effect), and \emph{\textbf{L}} meaning Low (having a Low multiplying effect). Examples include words like, \emph{``too"}, \emph{``more"} (\emph{H}); \emph{``hardly"}(\emph{N}); and \emph{``nearly"} (\emph{L}), etc.

\subsection{Location-Areas Set}
We also store the areas of about 20 major cities in India in \textbf{\emph{Location-Areas Set}}, to calculate the radii that are to be used during tweet extraction. Examples are New Delhi (1484 \emph{sq. km.}), Mumbai (603 \emph{sq. km.}), etc.

\section{Emotion-Detection Algorithm}
Our model consists of two completely different, yet interdependent approaches. The first approach uses Natural Language Processing, \emph{Emotion-Words Set} and several textual features. It attempts to classify and score text according to the emotions present in it. The second approach uses standard classifiers like SMO and J48 to classify tweets. Finally, we combine both these approaches to propose a Hybrid approach to detect emotions in text more effectively. Note that, even though we are extensively using the tweets example throughout this paper, this algorithm is very generic and can be used to detect and quantify emotions in any piece of text.

\subsection{Using NLP and EWS: First Approach}

The first approach uses the tweets in \emph{Tweets Set}, which are already free from any unwanted characters, hyperlinks or hashtags. Various \emph{Stanford CoreNLP} \cite{manning-EtAl} tools are used for a comprehensive linguistic analysis. The following steps comprise the first approach.

\subsubsection{Tokenizing and Annotating}
First, we use \emph{PTBTokenizer} \cite{tokenizer} to tokenize the tweets into sentences, which are further tokenized into tokens. Then, we remove all the stop words from these tokens. The filtered tokens are then annotated using the following \emph{CoreNLP annotators}:
%Now, the tokens of the tweets are pre-processed and annotated using several features.

%The same lemmatization is applied to the tweet words later.
%\subsubsection{Importing the EWS}
%The EWS is imported from the database as a hashmap and each of the words in this hashmap is mapped to an array of Emotion-Categories € {HAPPINESS, SADNESS, DISGUST, SURPRISE, ANGER, FEAR} along with their Intensity-Categories € {STRONG, MEDIUM, LIGHT}. 
%All words in this hashmap are lemmatized to their base form. Eg. ‘happiness’ and ‘happily’ are changed to ‘happy’. The same lemmatization is applied to the tweet words later.

%\subsubsection{Filtering and Annotating}

%\paragraph{NLP-Annotations}

\begin{itemize}
\item \emph{pos}: Parts of Speech (noun, verb, adjective, for example)
\item \emph{lemma}: Lemmatized version of that word.
\item \emph{ner}: For Named Entitiy Recognition.
\end{itemize}

We also fetch the Grammatical dependencies between the tokens in a sentence which include \emph{nsubj} (the subject of a sentence),  \emph{dObj} (the object of a sentence), \emph{advMod} (the relation between a word and an adverb), and \emph{neg} (the relation between a word and a negative word).

\subsubsection{Finding Hits}
In this step, the annotated (and filtered) tokens are matched against the words present in the \emph{EWS}. But first, all the words in the \emph{EWS} are lemmatized to their base form. Eg. \emph{``happiness"} and \emph{``happily"} are changed to \emph{``happy"}. While matching the tokens against the \emph{EWS}, only the tokens that are annotated as \emph{``O"} (the \emph{other} entity in Named Entity Recognition) are considered, because a named entity (location, time or a person word, for example) can never be an \emph{Emotion-word}. A matched token along with all its characteristics/annotations is stored as a \textbf{\emph{hit}}.

\subsubsection{Detecting person in context}
In our analysis, we aim at detecting the emotions expressed by the tweet's poster, stressing on the poster's feelings. Therefore, we try to differentiate between cases in which the emotions involved are in relation to the poster himself or someone else. For instance, the degree of sadness of the poster in the tweet \emph{"I am sad"} is much higher compared to the tweet, \emph{"He is sad"}, posted by the same person, as the former clearly and directly suggests the sadness of the poster himself. This is done by using the \emph{nsubj} annotation of the tokens, which tells us who the subject of the sentence is. After finding the subject of the sentence, we determine which \emph{person} (\emph{first}, \emph{second} or \emph{third}), that subject belongs to. For this, we maintain a list of \emph{first} (\emph{``I"}, \emph{``me"}, etc.), \emph{second} (\emph{``You"}, \emph{``your"}, etc.) and \emph{third person} (\emph{``He"}, \emph{``She"}, \emph{``They"}, etc.) pronouns. Also, if the \emph{nsubj} is a proper noun, then the \emph{person} of that \emph{nsubj} is considered to be \emph{third}. Next,  each \emph{hit} detected in the previous step is associated with a \emph{person}, by finding the nearest \emph{nsubj} to the \emph{hit} on its left side. If there is no \emph{nsubj} in a sentence, we associate the \emph{hit} with the \emph{first person}. For instance, the \emph{hit} \emph{``Nice"} in the tweet, \emph{``Nice to see you!!"} is associated to the \emph{first person}, as there is no \emph{nsubj} in the sentence.

%Next, each pronoun in the sentence is assigned a person (first, second or third), using pre-defined lists of first-person words (\emph{"I"}, \emph{"me"}, etc.), second-person words (\emph{"You"}, \emph{"your"}, etc.) and third-person words (\emph{"He"}, \emph{"She"}, \emph{"They"}, etc.). Now, each \emph{hit} detected is associated with a person, to identify whether the poster of the tweet is talking about himself/herself or someone else. This is important, in order to measure the degree of how much the poster is affected by whatever the tweet is about. 

\subsubsection{Effect of Negation and Degree-words}
There are many instances, when some words detected using the \emph{EWS} are actually used in an entirely opposite sense, because of negations like \emph{``not"}, \emph{``never"}, \emph{``don't"}, etc. For instance, the tweet \emph{``Not at all feeling excited for school!"} may result in getting \emph{Happiness} as its \emph{Emotion-Category} (because of the \emph{Emotion-word} \emph{``excited"}), if negations are not accounted for. Also, there are many words (mostly adverbs), that enhance the intensity of an emotion. Eg. \emph{``I feel good"} and \emph{``I feel too good"} contain the same \emph{Emotion-word} \emph{``good"} but the score given to the second sentence should be more. For detecting negations and degree-words, we search for tokens that have been annotated as \emph{neg} and \emph{advmod}, while determining the grammatical dependencies in step 1, and tag them as degree/negation words, if they lie in the \emph{Degree-Words Set}. The \emph{Degree-Intensity} of the degree-word associated (if any) with every hit is stored as an annotation to the hit.

\subsubsection{Emoticon Detection}
Emoticons are a very useful and informative source for detecting emotions in text. Since emoticons are direct ways of knowing the emotions of the user, they are given a heavy weightage, if found. This feature is very efficient and accurate, especially with the ever-rising popularity of emoticons. We have added a list of 100+ most commonly used emoticons in the \emph{EWS} to enhance our model, along with their \emph{Intensity-Categories} (\emph{STRONG} and \emph{MEDIUM}). The emoticon-detection process is done before tokenization and uses regex. The emoticons detected in a tweet are also treated as \emph{hits}.

\begin{table}[h]
\centering
\caption{Emotion-Scores of various feature combinations}
\label{emotScore}
\begin{tabular}{|l|l|l|}
\hline
\textit{\textbf{Intensity-Category}} & \textit{\textbf{Degree-Intensity}} & \textit{\textbf{emotScore}} \\ \hline
\textit{STRONG}                      & \textit{None}                      & 6                           \\ \hline
\textit{STRONG}                      & \textit{H}                         & 8                           \\ \hline
\textit{STRONG}                      & \textit{L}                         & 6                           \\ \hline
\textit{STRONG}                      & \textit{N}                         & 2*                           \\ \hline
\textit{MEDIUM}                      & \textit{None}                      & 4                           \\ \hline
\textit{MEDIUM}                      & \textit{H}                         & 6                           \\ \hline
\textit{MEDIUM}                      & \textit{L}                         & 6                           \\ \hline
\textit{MEDIUM}                      & \textit{N}                         & 4*                           \\ \hline
\textit{LIGHT}                       & \textit{None}                      & 2                           \\ \hline
\textit{LIGHT}                       & \textit{H}                         & 6                           \\ \hline
LIGHT                                & L                                  & 4                           \\ \hline
\textit{LIGHT}                       & \textit{N}                         & 4*                           \\ \hline
\end{tabular}
\end{table}

\begin{table}[h]
\centering
\caption{Emoticon Scores}
\label{emoticon_scores}
\begin{tabular}{|l|l|}
\hline
\textit{\textbf{Intensity-Category}} & \textit{\textbf{emotScore}} \\ \hline
\textit{STRONG}                      & 80                              \\ \hline
\textit{MEDIUM}                      & 40                              \\ \hline
\end{tabular}
\end{table}

\begin{table}[h]
\centering
\caption{Person Scores}
\label{persScore}
\begin{tabular}{|l|l|}
\hline
\textit{\textbf{person}} & \textit{\textbf{perScore}} \\ \hline
\textit{first}           & 10                         \\ \hline
\textit{second}          & 2                          \\ \hline
\textit{third}           & 1                          \\ \hline
\end{tabular}
\end{table}

\subsubsection{Scoring}
Every \emph{hit} is finally, a tuple of its features that contribute in the scoring process:

\begin{itemize}
\item \emph{lemma}: the lemmatized version of the hit word
\item \emph{Emotion-Category}: \emph{HAPPINESS, SADNESS, ANGER, SURPRISE, FEAR, DISGUST}
\item \emph{Intensity-Category}: \emph{STRONG, MEDIUM or LIGHT}
\item \emph{Degree-Intensity}: {\emph{H}, \emph{L}, \emph{N} or \emph{None}}
\item \emph{person}: \emph{first, second or third}
\end{itemize}

Based on the various possible combinations of these features, \textbf{\emph{emotScore}} and \textbf{\emph{perScore}} is calculated for each of the \emph{hits}. If the \emph{hit} is an English word (and not an emoticon), the values in Table \ref{emotScore} are used for calculating \emph{emotScore}, whereas for emoticons, Table \ref{emoticon_scores} is used. It should be noted that if the \emph{Degree-Intensity} of a hit is \emph{N} (negation effect), its \emph{Emotion-Category} changes. This means that \emph{HAPPINESS} turns into \emph{SADNESS}/\emph{ANGER} and vice-versa (with their \emph{emotScores} given in Table \ref{emotScore}, marked by asterisks), whereas the \emph{emotScores} of the remaining three \emph{Emotion-Categories} (\emph{DISGUST}, \emph{SURPRISE}, \emph{FEAR}) becomes zero, when the \emph{Degree-Intensity} is \emph{N}. Table \ref{persScore} is used for calculating the \emph{perScore} of a \emph{hit}. The final emotional score (\emph{Score}) of a tweet is a six-tuple, which is calculated by summing over all the \emph{hits} (of a particular \emph{Emotion-Category}) in all the sentences in the tweet for each of the six \emph{Emotion-Categories (\textbf{EC\textsubscript{j}}, j=1,2,...6)}. 

\begin{equation}
Score[EC\textsubscript{j}] = \sum_{}^{All\ Hits\ of\ EC\textsubscript{j}} (emotScore*perScore)
\end{equation}

\begin{equation}
RelScore[EC\textsubscript{j}] = \frac{Score[EC\textsubscript{j}]}{\sum\limits_{j=1}^{6} Score[EC\textsubscript{j}]} * 100
\end{equation}

Also, the relative score (\emph{RelScore}) is calculated to understand the percentage of each emotion in a tweet. If there are no \emph{Emotion-words} in a tweet, the value of scores of all \emph{Emotion-Categories} in \emph{Score} will be zero. For example, if the tweet has no words of the \emph{Emotion-Category FEAR}, then \emph{Score[FEAR]} is 0. An overview of the first approach is given in Fig \ref{first_approach}.

\subsection{Using Machine Learning: Second Approach}
We use another method for emotion-detection where we use a Machine Learning classifier. For this, the most important and critical step is to prepare a good training-set. We make use of the first approach to generate the training-set, which is free from any manual annotation and thus can be developed or updated quickly.

\begin{figure*}[]
  \begin{center}
  \includegraphics[scale=0.25]{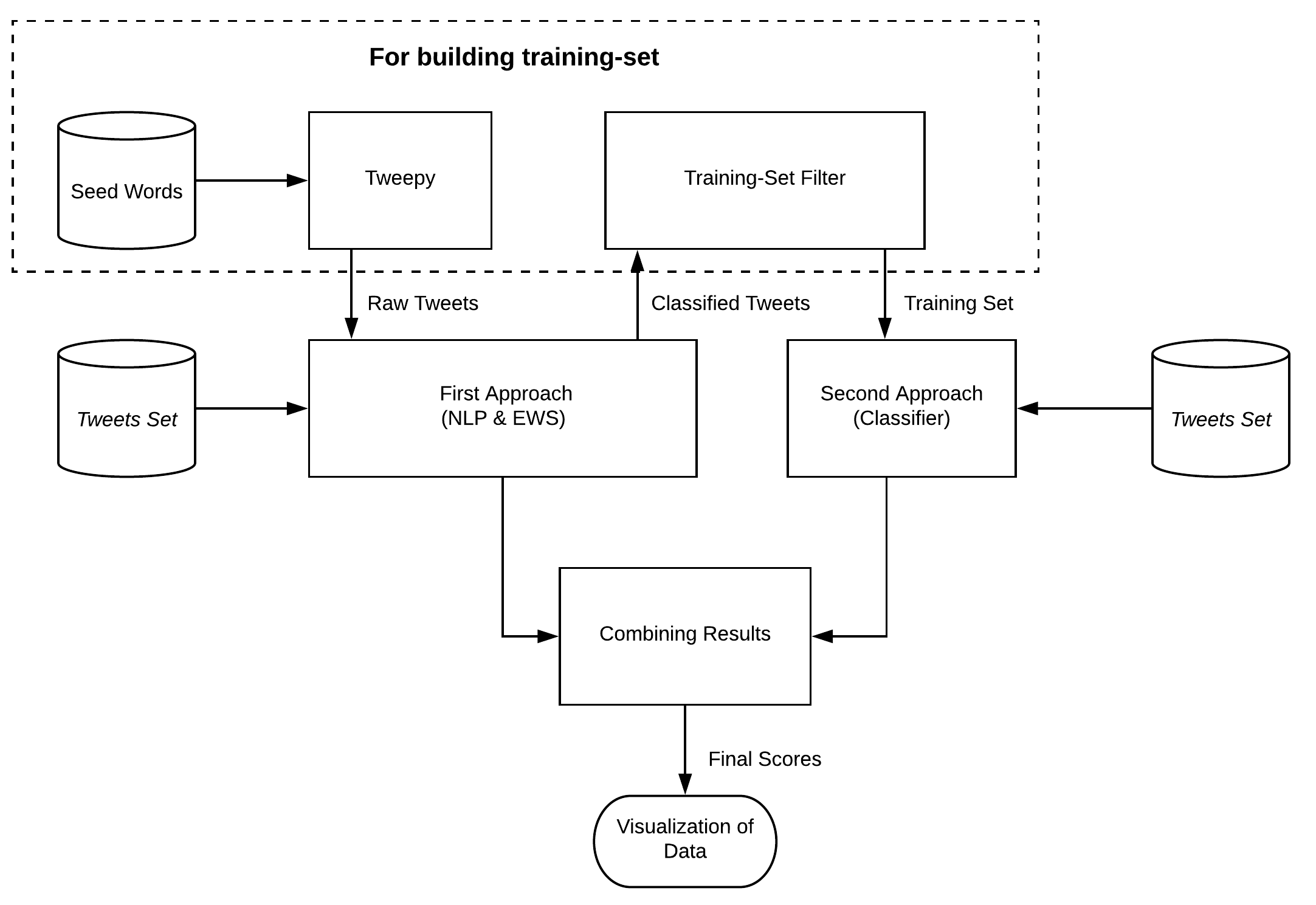}
  \caption{Combining the first and the second approach}
  \label{combine}
  \end{center}
\end{figure*}

\subsubsection{Generation of the training-set}
We have chosen a set of seed words, comprising of commonly used \emph{Emotion-words} and emoticons from the \emph{EWS}, evenly distributed over all the Emotion-Categories. We then query Tweepy using these seed words to develop a huge database of around 13,000 tweets. The seed words are used to ensure that we get tweets that express at least one of the six emotions. The retweets are removed to avoid any repetition of tweets. The even distribution of seed words ensures that we get an even distribution of tweets over all the Emotion-Categories, so that our classifier isn't biased. However, we have analysed that there is a large proportion of tweets on Twitter expressing happiness, so we have kept the number of seed words possessing the \emph{Emotion-Category HAPPINESS}, on the higher side. 
%Since, we already know which Emotion-Category each seed word belongs to, we already know the predicted Emotion-Category of the tweet.

\subsubsection{Filtering and Labelling the training-set}
For training an accurate classifier, we need all the tweets in the training-set to be strongly expressing only one of the six \emph{Emotion-Categories}. However, there are many tweets comprising of more than one emotion. Such type of tweets reduces the accuracy of the classifier, as we can give only one label to a tweet in the training set. If a tweet contains more than one emotion in the training-set, it will result in faulty learning as words related to the other (unlabeled) \emph{Emotion-Category} will also be treated as those related to the category labeled. Thus, we use the first approach to label all the 13,000 tweets and filter out the tweets having none or mixed emotions. Only those tweets which have a percentage of more than 70\% for a particular emotion are labeled and fed to the classifier for training. The final distribution of tweets in our labeled training set is as shown in Table \ref{training_set}.

\begin{table}[h]
\centering
\caption{Distribution of tweets in the training-set}
\label{training_set}
\begin{tabular}{|l|l|}
\hline
\textit{\textbf{Emoticon-Category}} & \textbf{No. of tweets} \\ \hline
HAPPINESS                           & 2617                   \\ \hline
SADNESS                             & 1416                   \\ \hline
\textit{FEAR}                       & 1459                   \\ \hline
\textit{DISGUST}                    & 776                    \\ \hline
\textit{ANGER}                      & 1316                   \\ \hline
\textit{SURPRISE}                   & 944                    \\ \hline
\textbf{Total}                   & \textbf{8528}                    \\ \hline
\end{tabular}
\end{table}

\subsubsection{Training the classifiers}
We use the open source library Weka \cite{weka} for the implementation of the classifiers. We use two very popular Weka classifiers, SMO \cite{smo} and J48 \cite{j48}. Before classification, we applied the following pre-processing steps on the data:

\begin{itemize}
\item Stop-Word Filtering
\item Lower-casing all words
\item Stemming each word using Weka's Snowball Stemmer
\end{itemize}

The classifiers output the relative probabilities of each of the six \emph{Emotion-Categories}, for a tweet (or a piece of text). The \emph{Emotion-Category} with the highest probability is called the \textbf{\emph{Labeled-Category (L\textsubscript{c})}}.

\subsection{Combining First and Second Approach Results}

We finally combine the first and the second approach. The scores which are generated by the first approach are modified, according to the \emph{Labeled-Category} of the classifier. The score of only the \emph{Labeled-Category} is modified as follows:

\begin{equation}
FinalScore[L\textsubscript{c}] = Score[L\textsubscript{c}] + (0.2*Score[M\textsubscript{c}])
\end{equation}

where $M_c$ refers to the label with the maximum score, after the first approach results. This is done to give weightage to the results of the classifier by increasing the score of the classified category in proportion to the maximum so as to avoid too little or too much relative change in the scores. This will definitely make a difference in deciding the final \emph{Emotion-Category}, when two different \emph{Emotion-Categories} are too close by or equal and are in competition. After this, the \emph{Emotion-Category} with the maximum final score is decided as the final \emph{Emotion-Category} of the tweet (or the piece of text). An overview of the combined approach is shown in Fig \ref{combine}.

\section{Results And Analysis}

\subsection{Testing Results}
Similar to the training-set, we have created a testing-set of tweets by extracting tweets using some seed words and then using the first approach to filter and label these emotional tweets. This set consists of a total of 900 tweets, where each \emph{Emotion-Category} has around 150 tweets, so as to maintain uniformity. Also, it is ensured that all tweets in the testing set are different from those in the training set. The two chosen classifiers, on testing with the testing-set, gave the results as shown in Table \ref{smo} and \ref{j48}. The correctly classified instances are the tweets for which the expected \emph{Emotion-Category} matches the actual \emph{Emotion-Category}. As can clearly be seen from the tables, we have achieved a remarkable accuracy of 91.7\% for SMO and 85.4\% for J48, which proves the merits of our Emotion-Detection Algorithm.

\begin{table}[h]
\centering
\caption{Accuracy of the SMO Classifier}
\label{smo}
\begin{tabular}{|l|l|l|}
\hline
\multicolumn{3}{|c|}{\textbf{SMO}}                          \\ \hline
Correctly Classified Instances   & 826   & \textbf{91.7\%}  \\ \hline
Incorrectly Classified Instances & 74    & 8.22\%           \\ \hline
Total No. of Instances           & \multicolumn{2}{l|}{900} \\ \hline
\end{tabular}
\end{table}

\begin{table}[h]
\centering
\caption{Accuracy of the J48 Classifier}
\label{j48}
\begin{tabular}{|l|l|l|}
\hline
\multicolumn{3}{|c|}{\textbf{J48}}                          \\ \hline
Correctly Classified Instances   & 769   & \textbf{85.4\%}  \\ \hline
Incorrectly Classified Instances & 131   & 14.5\%           \\ \hline
Total No. of Instances           & \multicolumn{2}{l|}{900} \\ \hline
\end{tabular}
\end{table}

\subsection{Surety Factor ($S_f$)}
The surety factor indicates how confident and assertive our analysis and results are, on a given text/tweet. Its value is low, when there is a mismatch between the results of the two approaches or when the text seems to lack any emotions. On the other hand, its value is high when the two approaches concur with each other in results, there are too many \emph{hits} or when one of the emotion-scores is very high. The surety factor is calculated on a scale of 6 and is dependent on several factors:

\begin{itemize}
\item Classifier\_label\_match: Whether the classifier label matches the category of the maximum score of the first approach. (Value: True/False).
\item Max\_score: The value of the maximum of all the six scores.
\item Max\_percent: The relative percentage of the Max\_score over all six scores.
\item Second\_diff: The difference between the maximum and second maximum score value as a percent of the maximum.
\item Hits: The number of \emph{Emotion-words} matched in the text.
\end{itemize}

Surety factor is calculated differently for the following two cases:

\begin{itemize}
\item If the \emph{hits} in the tweet/text belong to the same \emph{Emotion-Category}, then only the factors Classifier\_label\_match and Max\_score are used.
\item If the \emph{hits} in the tweet/text belong to different \emph{Emotion-Categories}, all five factors are considered.
\end{itemize}

\subsection{Visualization of Results}
\begin{figure*}[]
  \includegraphics[width=\linewidth]{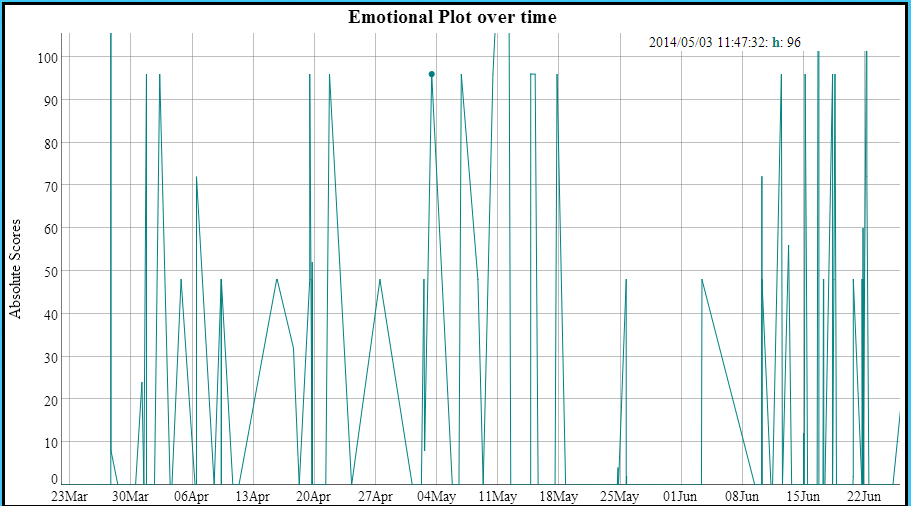}
  \caption{Time varying happiness plot of a twitter-user}
  \label{swings}
\end{figure*}

\begin{figure*}[]
  \includegraphics[width=\linewidth]{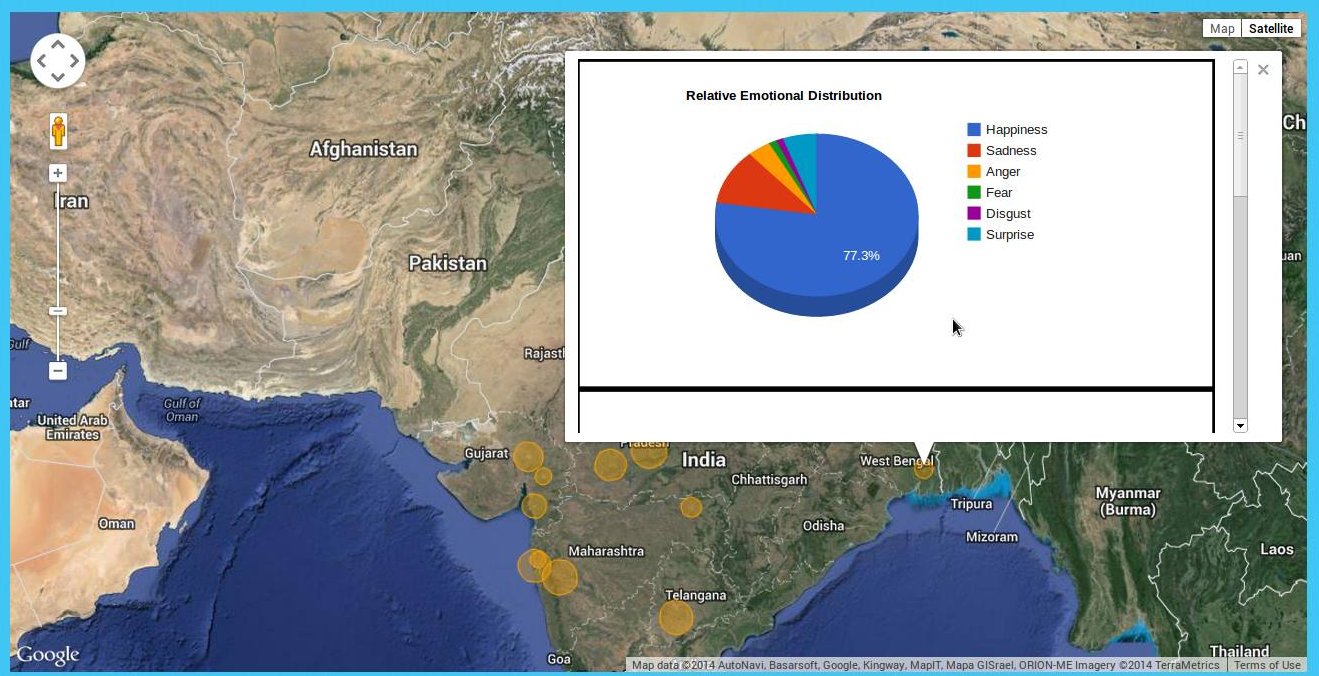}
  \caption{Location-wise emotion-analysis of major cities in India}
  \label{location}
\end{figure*}

\begin{figure*}[]
  \includegraphics[width=\linewidth]{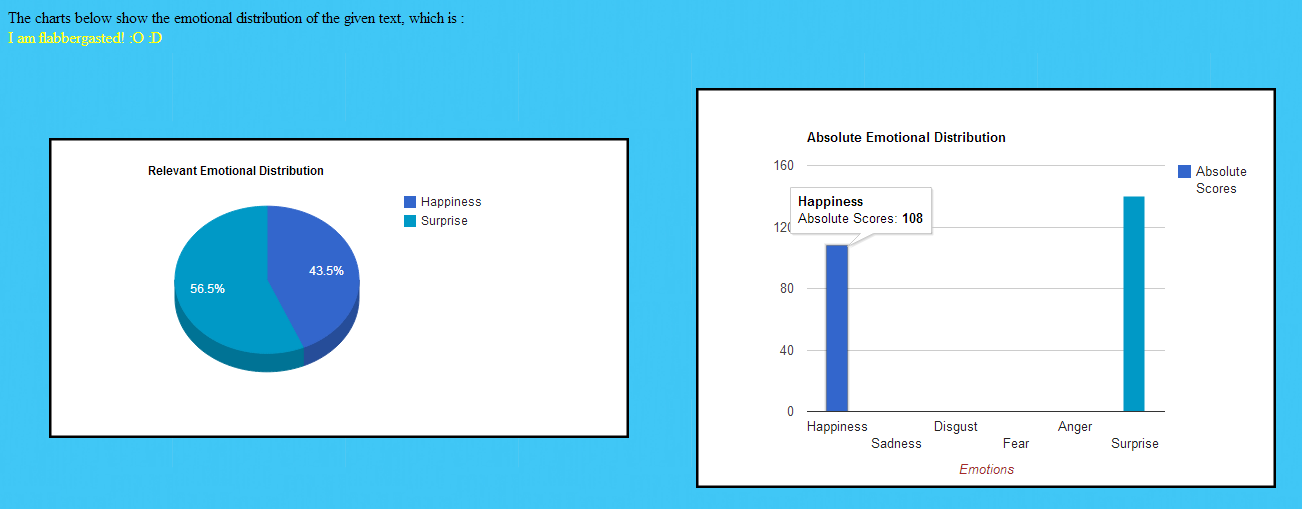}
  \caption{Detecting emotions in a piece of text}
  \label{piechart}
\end{figure*}

In order to demonstrate the usefulness of the proposed Emotion-Detection Algorithm, we have implemented the following applications: 

\begin{itemize}
\item One-user Analysis (twitter account): We have developed an interface, which takes as input, the username of any twitter-user, processes all his/her tweets till date and displays a time-varying mood-swings plot as well as the relative and absolute emotional distribution in the form of pie charts. Fig. \ref{swings} is an example of the varying happiness of a twitter-user over time.
\item Location Analysis: Using Google Maps API, a world map is displayed and each circle on it represents the emotional analysis of that particular region. The radius of each circle is proportional to the area of the region. This analysis is done for around 20 major cities in India, using the \emph{Location-Areas Set} (See Fig. \ref{location})
\item Document Analysis: Another interface takes as input, entire documents or blocks of text, processes it, and displays the relative and absolute emotional distribution of the document in the form of pie charts. (See Fig. \ref{piechart} for an example sentence.)
\end{itemize}

%\begin{figure}
%\centering
%\begin{minipage}{.5\textwidth}
%  \centering
%  \includegraphics[width=.4\linewidth]{photo2.png}
%  \captionof{figure}{A figure}
%  \label{fig:test1}
%\end{minipage}%
%\begin{minipage}{.5\textwidth}
%  \centering
%  \includegraphics[width=.4\linewidth]{photo2.png}
%  \captionof{figure}{Another figure}
%  \label{fig:test2}
%\end{minipage}
%\end{figure}

\section{Conclusion}
In this paper, we have addressed the problem of classifying text into the six basic \emph{Emotion-Categories}, rather than just labeling them as positive or negative. Through our research and a self-generated reliable bag of emotional words (\emph{EWS}), we can now effectively quantify various emotions in any block of text. We have also automatically generated a labeled training-set (without manually labeling the tweets) of emotionally-biased tweets using a keyword-matching approach, which was then used to train various classifiers. Moreover, we have also introduced the concept of Surety Factor to suggest the reliability of our output and the degree of usefulness and correctness of our results. Finally, we visualized our results using pie-charts, bar-graphs and maps, and demonstrated the various applications of our analysis. In future, a system could be established for automatically updating the bag-of-words which we created, on the basis of new tweets and data analysed. Using our approach, many interesting apps can be created, such as an add-on to a social-networking site displaying the recent mood of each of your friends. Also, our analysis of Twitter can be extended to the development of a real-time system, analyzing mood-swings and emotions on Twitter.

\bibliographystyle{unsrt} 
\bibliography{mybib}

%\begin{thebibliography}{00}
%\bibitem{b1} G. Eason, B. Noble, and I. N. Sneddon, ``On certain integrals of Lipschitz-Hankel type involving products of Bessel functions,'' Phil. Trans. Roy. Soc. London, vol. A247, pp. 529--551, April 1955.
%\bibitem{b2} J. Clerk Maxwell, A Treatise on Electricity and Magnetism, 3rd ed., vol. 2. Oxford: Clarendon, 1892, pp.68--73.
%\bibitem{b3} I. S. Jacobs and C. P. Bean, ``Fine particles, thin films and exchange anisotropy,'' in Magnetism, vol. III, G. T. Rado and H. Suhl, Eds. New York: Academic, 1963, pp. 271--350.
%\bibitem{b4} K. Elissa, ``Title of paper if known,'' unpublished.
%\bibitem{b5} R. Nicole, ``Title of paper with only first word capitalized,'' J. Name Stand. Abbrev., in press.
%\bibitem{b6} Y. Yorozu, M. Hirano, K. Oka, and Y. Tagawa, ``Electron spectroscopy studies on magneto-optical media and plastic substrate interface,'' IEEE Transl. J. Magn. Japan, vol. 2, pp. 740--741, August 1987 [Digests 9th Annual Conf. Magnetics Japan, p. 301, 1982].
%\bibitem{b7} M. Young, The Technical Writer's Handbook. Mill Valley, CA: University Science, 1989.
%\end{thebibliography}

\end{document}